\documentclass[aps,pra]{revtex4}
\usepackage{graphicx}
\usepackage{amsmath}
\usepackage{amssymb}

\begin{document}

\title{Shared quantum control via sharing operation on remote single qutrit}
\author{Daochu Liu$^a$, Yimin Liu$^b$, Xiaofeng Yin$^a$, Xiansong Liu$^a$, Zhanjun Zhang$^{a,\dag}$ \\
{\normalsize $^a$ School of Physics \& Material Science, Anhui
University, Hefei 230039, Anhui, China} \\
{\normalsize $^b$ Department of Physics, Shaoguan University, Shaoguan 512005, China} \\
{\normalsize $^{\dag}$zjzhang@ahu.edu.cn } }

\maketitle

\begin{minipage}{420pt}
{\bf Abstract} Two qubit operation sharing schemes [{\it J. Phys.} B {\bf 44} (2011) 165508]
are generalized to qutrit ones. Operations to be shared are classified into three different classes
in terms of different probabilities (i.e, 1/3, 2/3 and 1). For the latter two
classes, ten and three restricted sets of operations are found out, respectively. Moreover,
the two generalized schemes are amply compared from four aspects, namely, quantum and classical
resource consumption, necessary-operation complexity, success probability and efficiency.
It is found that the second scheme is overall more optimal than the first one as far as
three restricted sets of operations are concerned.


\end{minipage}\\\\

\noindent {\bf 1 \ Introduction}

Entanglement is admitted as a kind of important quantum resource nowadays.
In the last two decades it has been extensively exploited and utilized
in the many fields of quantum information science to fulfill various quantum
tasks involving classical information processing in a quantum manner and
quantum information (i.e., quantum state) processing[1-25], such as quantum key distribution,
quantum state teleportation, remote state preparation, operation teleportation,
state sharing, direct secure communication, quantum computing, and so on.
Enlightened by the generalization of quantum state teleportation to
quantum state sharing, in 2011 Zhang and Cheung[26] definitely presented
quantum operation sharing with shared entanglements.
They introduced the sharing idea into quantum operation teleportation
and proposed two specific schemes with different groups of shared entanglements.
The first one is a universal but nontrivial scheme for sharing any single-qubit
operation with one group of shared entanglements, while the latter treats
the sharing of two restricted sets of operations with less resource consumptions.
Incidentally, these schemes are uniformly referred to as the ZC schemes later.
As same as quantum operation teleportation, QOS can be viewed as a remote control
(encryption, decryption or destruction) on quantum information in the future quantum network.
If the target state as an important quantum information is initially
encrypted with a given unitary operation, then the inverse operation on it is
obviously the decryption on it. Moreover, it is natural to regard a random operation
as the destruction on it. As a consequence, quantum operation sharing can be taken
as a key to activate some important actions in the future lives by some sharer
entities, such as missile emissions, quantum collective seal or unseam, remote
joint destruction of quantum money, and so on. Therefore, recently this topic
has already attracted some attentions[27-29]. In spite of this, it is worthy emphasizing
that all these works treat only the sharing of single-qubit operations with shared qubit entanglements.

In the intending quantum networks, various quantum states involving high-dimensional
qudit state might be employed and distributed among different nodes due to some special
demands, such as peculiar quantum tasks in some concrete quantum scenarios, some definite
security requirements, and so on. Because of this, some researchers
have been attracted by the issues related to high-dimensional qudit cases
and started to explore them[30-39]. Surely, their works includes
the employment of qutrit entanglements in addressing various relative quantum problems[38].
In this condition, it is intriguing to ask what will happen in the field of QOS if
the accessible quantum channels are composed of shared qutrit entanglements?
Specifically, what kind of qutrit states can be used to
fulfill QOS? If quantum channels are determined, what is the maximal success probability for
sharing a given operation? Can the operation be shared more economically with less operation complexity?
Can operations be classified with different classes corresponding to different
success sharing probabilities? How to characterize those classes and find out them?
and so on and so forth. To our best knowledge these mentioned issues have not been touched by far.
Hence, it is of interest to consider the generalization of the ZC schemes[26] from
the aspect of particle degree. Since there exist many open questions,
in this paper we will consider a comparatively simple generalization, i.e.,
generalize the simplest QOS schemes (i.e., the ZC schemes) to the qutrit ones
by employing the qutrit Bell and GHZ states as quantum channels.
Such extensions is actually a little intricate but will lead to quite abundant results.
More importantly, they indicate that QOS as remote control can be  achieved with
shared qutrit entanglement, too. We will show them later.
Additionally, at present it is broadly recognized that resource consumption and operation complexity
as hot topics in many fields are always concentrated. How to consume less amount of resources and how to
degrade the difficulty and intensity of necessary operations are continuously
attracting much attention and pursued by many researchers. Because of this,
in this paper we will focus our attention on this issue during the generalizations, too.

The rest of this paper is organized as follows. In section 2, two three-party schemes
for sharing single-qutrit operations on a remote qutrit in any state are preciously
proposed with different quantum channels. In section 3, two schemes are amply compared
and discussed from the four aspects of quantum and classical resource consumption,
necessary-operation complexity, success probability and efficiency. Finally,
a concise summary is made in section 4.  \\

\noindent {\bf 2 \ Remote sharing of single-qutrit operation}

Now let us start to present our schemes. In either scheme there are three legitimate users,
say, Alice, Bob and Charlie. Alice is the initial performer of a single-qutrit operation ${\cal U}$.
Incidentally, whether she knows it or not is uncertain and will be discussed and treated separably later.
Bob and Charlie are Alice's two remote agents. Alice needs to perform the operation ${\cal U}$
on a qutrit in state $|\chi \rangle$ at one remote agent's site. She wants to fulfill the task
with her agents' assistance and by making full use of the quantum and classical channels
linking the three legitimate users. However, she trusts neither agent but their entity. Specifically,
she should assure that the operation can not be successfully executed on the qutirt
by either agent solely but conclusively achieved via the mutual collaboration of her two
agents. Moreover, the state $|\chi \rangle$ in the qutrit can be arbitrary.
Suppose the state $|\chi \rangle$ is initially in Bob's qutrit $b''$ and can be written as
\begin{eqnarray}
|\chi\rangle_{b^{''}}=\alpha|0\rangle_{b''}+\beta|1\rangle_{b''}+\gamma|2\rangle_{b''},
\end{eqnarray}
where $ \alpha $, $\beta$ and $\gamma$ are complex and satisfy $|\alpha{|}^2+|\beta|^2+|\gamma|^2=1$.

\vskip 0.2cm
\noindent {\bf 2.1 General scheme for arbitrary single-qutrit operation}

Now let us put forward our general three-party QOS scheme, which is universally
applicable for sharing an arbitrary single-qutrit operation.
The schematic demonstration is illustrated in figure 1. The scheme can be concisely depicted as follows.

\begin{figure}[h]
\begin{center}
\includegraphics[width=6in]{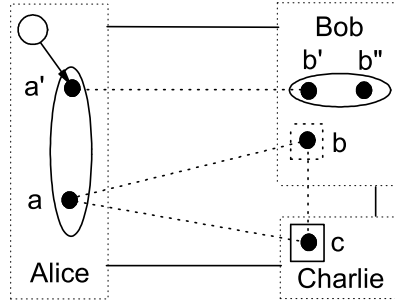}
\vskip -4cm
\begin{minipage}{330pt}
\caption{ Illustration of our general three-party QOS scheme with generalized GHZ and generalized Bell states.
Dotted rectangles are participants' locations. Solid lines among rectangles stand for classical communication
channels. Solid dots denote qutrits. Dotted lines linking qutrits means entanglements.
The solid circle labels the unitary operation ${\cal U}$ to be shared. Solid ellipses
represent generalized Bell-state measurements. The solid and dotted squares illustrate
the single-qutrit measurement and unitary operation, respectively. See text for more details.}
\end{minipage}
\end{center}
\end{figure}

(i) {\it Initial stage}. \ In this scheme, the quantum channels linking the three legitimate users are a shared generalized Bell state $|{\cal B}_{00}\rangle$
and a shared generalized GHZ state $|{\cal G}\rangle$, i.e.,
\begin{eqnarray}
|{\cal B}_{00}\rangle_{a'b'}=\frac{1}{\sqrt3}(|00\rangle+|11\rangle+|22\rangle)_{a'b'},\\
|{\cal G}\rangle_{abc}=\frac{1}{\sqrt3}(|000\rangle+|111\rangle+|222\rangle)_{abc},                                                                                                                        \end{eqnarray}
where the qutrit trio $(b,b',b'')$ belongs to Bob, the qutrit pair $(a,a')$ to Alice and the qutrit $c$ to Charlie.

(ii) {\it QT process}. \   In this stage, the state $|\chi\rangle$ in Bob's qutrit $b''$ is teleported to Alice's qutirt $a'$ via the
standard QT process with the generalized Bell state $|{\cal B}_{00}\rangle$ as the quantum channel between Alice and Bob. Alternatively,
after the process the state of qutrit $b''$ has been swapped to the qutrit $a'$ and hence the state of qutrit $a'$ is transformed to $|\chi\rangle$.

(iii) {\it Operation performance}. \   Alice carries out the operation ${\cal U}$ on her qutrit $a'$ in state $|\chi\rangle$, i.e.,
 \begin{eqnarray}
{\cal U} \rightarrow |\chi\rangle_{a'} \Longrightarrow ({\cal U} |\chi\rangle)_{a'}.
\end{eqnarray}

(iv) {\it QSTS process}. \
In this stage, the state ${\cal U} |\chi\rangle$ in Alice's qutrit $a'$ is shared by Bob and Charlie
via a standard QSTS process with the generalized GHZ state $|{\cal G}\rangle_{abc}$ as the quantum channel.
In other words, if Bob and Charlie collaborate with each other, they can finally reconstruct the state
${\cal U} |\chi\rangle$ on either the qutrit $b$ or the qutrit $c$. This result is actually equivalent to
Alice's aim, that is, conclusively performing her single-qutrit operation ${\cal U}$ on a remote qutrit
in state $\chi$ at an agent's position.

\vskip 0.2cm
\noindent {\bf 2.2 Specific scheme for restricted sets of operations}

Now let us present another tripartite QOS scheme with quantum channels
different from those in the first scheme. The illustration of this scheme is shown in figure 2.
The details of the scheme are described as follows.

(I) {\it Initial stage}. \   The present quantum channels linking the three legitimate users are
\begin{eqnarray}
|{\cal B}_{00}\rangle_{a'b'}&=&\frac{1}{\sqrt3}(|00\rangle+|11\rangle+|22\rangle)_{a'b'},\\
|{\cal B}_{00}\rangle_{ac}&=&\frac{1}{\sqrt3}(|00\rangle+|11\rangle+|22\rangle)_{ac},                                                                                                                       \end{eqnarray}
where the qutrit pair $(a,a')$ is at Alice's hand, and the qutrits $b'$ and $c$ are in Bob's and Charlie's sites, respectively.
Note that, different from the quantum channels employed in the general scheme,
here a generalized Bell state $|{\cal B}_{00}\rangle_{ac}$ is used
instead of the generalized GHZ state $|{\cal G}\rangle_{abc}$ there.

\begin{figure}[h]
\begin{center}
\includegraphics[width=6.0in]{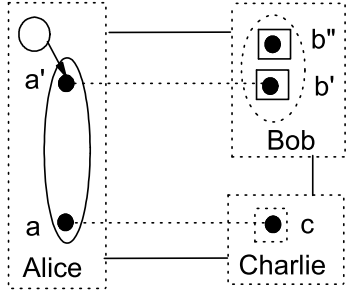}
\vskip -4.5cm
\begin{minipage}{300pt}
\caption{ Illustration of our specific three-party QOS scheme with two generalized Bell states.
The same as fig.1 with the dotted ellipse representing the unitary operation ${\cal V}$.
See text for more details.}
\end{minipage}
\end{center}
\end{figure}

\vskip 0.2cm
(II) {\it Bob's performances.} \ \ First, Bob performs a unitary operation ${\cal V}$ on his qutrit
pair $(b',b'')$, where
\begin{eqnarray}
{\cal V} &=&|00\rangle\langle00|+|01\rangle\langle01|+|02\rangle\langle02|+|10\rangle\langle11|
+|11\rangle\langle12|\nonumber\\
&+&|12\rangle\langle10|+|20\rangle\langle22|+|21\rangle\langle20|+|22\rangle\langle21|.
\end{eqnarray}
After the operation, the state of the qutrit trio $(a', b', b'')$ is converted into
\begin{eqnarray}
{\cal V}_{b'b''}|\Psi_{00}\rangle_{a'b'}|\chi\rangle_{b''}
&=&  \{ [\alpha |00\rangle + \beta |11\rangle + \gamma |22\rangle ]_{a'b'} |0\rangle_{b''}
+ [\alpha |22\rangle + \beta |00\rangle + \gamma |11\rangle ]_{a'b'} |1\rangle_{b''}
\nonumber\\ &+& [\alpha |11\rangle + \beta |22\rangle
+ \gamma |00\rangle ]_{a'b'} |2\rangle_{b''} \}/{\sqrt 3}.
\end{eqnarray}
Then Bob measures his qutrit $b''$ in the bases $\{|0\rangle,|1\rangle,|2\rangle\}$.
If Bob measures $|0\rangle_{b''}$, then he does the identity operation (i.e., nothing)
on his qutrit $b'$. Otherwise, if he gets $|1\rangle_{b''}$ ($|2\rangle_{b''}$),
he executes the unitary operation $S=|0\rangle\langle 2|+|2\rangle\langle 1|+|1\rangle\langle 0|$
($T=|0\rangle\langle 1|+|2\rangle\langle 0|+|1\rangle\langle 2|$) on his qutrit $b'$.
In terms of their prior agreement that the single-qutrit state $|i\rangle$ corresponds
to the classical single trit $i$ and vice versa (same hereafter), Bob
notifies Charlie of the measurement result via the classical channel linking them.

\vskip 0.2cm
(III) {\it Alice's performances.} \ \ Upon receiving Bob's message, Alice first does
the same single-qutrit operation as Bob's on her qutrit $a'$.
Bob's and Alice's performances lead to one of the following transformations:
\begin{eqnarray}
&&[\alpha |00\rangle + \beta |11\rangle + \gamma |22\rangle ]_{a'b'} |0\rangle_{b''}
\xrightarrow {\ I_{a'} I_{b'} }
[\alpha |00\rangle + \beta |11\rangle + \gamma |22\rangle ]_{a'b'} |0\rangle_{b''},  \\
&&[\alpha |22\rangle + \beta |00\rangle + \gamma |11\rangle ]_{a'b'} |1\rangle_{b''}
\xrightarrow {S_{a'}S_{b'}}
[\alpha |00\rangle + \beta |11\rangle + \gamma |22\rangle ]_{a'b'} |1\rangle_{b''}, \\
&&[\alpha |11\rangle + \beta |22\rangle + \gamma |00\rangle ]_{a'b'} |2\rangle_{b''}
\xrightarrow {T_{a'}T_{b'}}
[\alpha |00\rangle + \beta |11\rangle + \gamma |22\rangle ]_{a'b'} |2\rangle_{b''}.
\end{eqnarray}
Note that in the formulae above the operations on qutrits $a'$ and $b'$ are same.
Obviously, one can see that the state of the qutrit pair $(a', b')$ is always $[\alpha |00\rangle + \beta |11\rangle + \gamma |22\rangle ]_{a'b'}$
after their operations. Afterwards, she performs the unitary operation ${\cal U}$ on her qutrit $a'$. In this case the
state of Alice's qutrit pair becomes
\begin{eqnarray}
|J\rangle_{a'b'}= {\cal U}_{a'}[\alpha |00\rangle + \beta |11\rangle + \gamma |22\rangle ]_{a'b'}.
\end{eqnarray}
At this moment the state of the qutrit quadruple ($a', a, b', c$) is
\begin{eqnarray}
|Q\rangle_{a'ab'c}= |J\rangle_{a'b'} \otimes |{\cal B}_{00}\rangle_{ac}.
\end{eqnarray}
Subsequently, Alice measures her qutrit pair with the generalized Bell states as the measuring bases
\begin{eqnarray}
\{ |\Psi_{n,m}\rangle=\sum\limits_{j=0}^2 e^{\frac{2nj \pi}{3} i} |j\rangle\otimes|(j+m)~\textrm{mod}~ 3\rangle/\sqrt{3}; \ \ n\in{(0,1,2)}, \ m\in{(0,1,2)} \}.
\end{eqnarray}
After Alice's measurements, the state of the qutrit quadruple has collapsed to one of the following nine states:
\begin{eqnarray}
|\Psi_{n,m}\rangle_{a'a}\langle\Psi_{n,m}|Q\rangle&=& \frac{1}{3}|\Psi_{n,m}\rangle_{a'a}\sigma^{(n,m)}_c
{\cal U}_{c}(\alpha|00\rangle+\beta|11\rangle+\gamma|22\rangle)_{cb'}, \ \  n,m=0,1,2,
\end{eqnarray}
where
\begin{eqnarray}
\sigma^{(n,m)} = |m\rangle\langle0| + e^{\frac{4n\pi}{3} i }|(m+1)~\textrm{mod}~3\rangle\langle1|
+ e^{\frac{2n\pi}{3} i }|(m+2)~\textrm{mod}~3\rangle\langle2|.
\end{eqnarray}
Then at last, she publishes via the classical channels the outcome
according to the prior agreement,  i.e., the generalized Bell state
$|\Psi_{n,m}\rangle$ corresponds to the classical trit pair $(n,m)$
and vice versa (same hereafter).
\vskip 0.2cm

(IV) {\it Reconstruction via agents' collaboration.}

If Bob and Charlie collaborate with each other, they can deterministically
or probabilistically reconstruct the operation ${\cal U}$ on the state
$|\chi\rangle$ in either qutrit $c$ or qutrit $b'$.
Without loss of generality, suppose they decide to restore it in qutrit $c$.
In this condition, Charlie starts to perform a unitary operation
$\sigma^{(n,m)\dagger}$ on his qutrit $c$ after Alice publishes the outcome.
His operation makes the qutrit pair $(b',c)$ in the state
\begin{eqnarray}
|J\rangle_{cb'}= {\cal U}_{c}[\alpha |00\rangle + \beta |11\rangle + \gamma |22\rangle ]_{cb'}.
\end{eqnarray}
Subsequently, Bob measures his qutrit $b'$ with the orthonormal measuring bases defined as
\begin{eqnarray}
 |\xi_{0}\rangle = |\vec{V}(1/{\sqrt 3},1/{\sqrt 3},1/{\sqrt 3})\rangle,  \ \
|\xi_{1}\rangle =  e^{i\tau_1}|\vec{V}_1(x_1,y_1,z_1)\rangle,\ \ |\xi_{2}\rangle = e^{i\tau_2}|\vec{V}_2(x_2,y_2,x_2)\rangle,
\end{eqnarray}
where $x$'s, $y$'s and $z$'s are complex, $\tau$'s are arbitrarily real, and
$|\vec{V}_i(x,y,z)\rangle \equiv x|0\rangle+y|1\rangle+z|2\rangle$.
The orthogonality and normality of the measuring bases require that
\begin{eqnarray}
&& z_1=-x_1-y_1, |x_1|^2+|y_1|^2+|x_1+y_1|^2=1, \nonumber\\
&& x_2 = x'_2/N =  (-1+2x_{1}x_{1}^{\ast}+x_{1}y_{1}^{\ast})/N, \nonumber\\
&& y_2 = y'_2/N =  (2y_{1}x_{1}^{\ast}+y_{1}y_{1}^{\ast})/N, \nonumber\\
&& z_2 = z'_2/N = (1-2x_{1}x_{1}^{\ast}-2y_{1}x_{1}^{\ast}-x_{1}y_{1}^{\ast}-y_{1}y_{1}^{\ast})/N, \nonumber\\
&& N=\sqrt{|x'_2|^{2}+|y'_2|^{2}+|z'_2|^{2}}.\nonumber
\end{eqnarray}
It is obvious that the first basis is unchanged as a fixed vector
in the three-dimensional Hilbert space, while the latter two are
variant as the function of three dependent parameters. Nonetheless,
here it is necessary to stress that parameters $x_1$ and $y_1$ are
not completely independent but correlated to each other by a constraint.
By virtue of the measuring bases, the state of the qutrit pair $(c,b')$
can be rewritten as
\begin{eqnarray}
|J\rangle_{cb'} &=& \frac{1}{\sqrt{3}}\{[{\cal U}|\chi\rangle]_{c}|\xi_{0}\rangle_{b'}
+[{\cal U} W_1(x_1,y_1,z_1)|\chi\rangle]_{c}|\xi_1\rangle_{b'} \nonumber \\
&+& [{\cal U} W_2(x_2,y_2,x_2)|\chi\rangle]_{c}|\xi_2\rangle_{b'}\},
\end{eqnarray}
where
\begin{eqnarray}
W_k(x,y,z)=e^{-i\tau_k} (x^*|0\rangle\langle0|+y^*|1\rangle\langle1|+z^*|2\rangle\langle2|), \ \ \ \ k=1,2.
\end{eqnarray}
Incidentally, it is obvious that $W_k(x,y,z)$ is unitary. Consequently,
if Bob measures $|\xi_{0}\rangle_{b'}$, Alice's operation ${\cal U}$
has been conclusively performed on Charlie's qutrit $c$. However,
$|\xi_{0}\rangle_{b'}$ occurs only with probability 1/3. It is quite possible
that Bob measures $|\xi_1\rangle_{b'}$ or $|\xi_2\rangle_{b'}$.
In these two cases, intuitively, one is readily to see that the operation ${\cal U}$
has not been successfully executed on the qutrit $c$. Whether Alice's goal
can be achieved in the end is still uncertain and completely determined
by the relations among ${\cal U}$ and $W_1(x_1,y_1,z_1)$ as well as $W_2(x_2,y_2,x_2)$.
With probability 1/3 Bob may measure $|\xi_1\rangle_{b'}$. In this case,
if ${\cal U}W_1(x_1,y_1,z_1)=\pm {\cal U}W_1(x_1,y_1,z_1)$
holds, then based on Bob's message about the measurement result,
Charlie can achieve Alice's goal by executing the unitary operation $W_1^\dag(x_1,y_1,z_1)$
on his qutrit $c$. Otherwise, Alice's goal can not be achieved.
Similarly, Bob may measure $|\xi_2\rangle_{b'}$ with probability 1/3, too.
In this case, if ${\cal U}W_2(x_2,y_2,x_2)=\pm {\cal U}W_2(x_2,y_2,x_2)$ holds,
then Charlie can fulfill Alice's operation on his qutrit $c$ by performing
the reversal unitary operation of $W_2(x_2,y_2,x_2)$ in terms of Bob's message
on the measurement result. Otherwise, Alice's operation on a remote qutrit
in agents' site fails. Surely, if both ${\cal U}W_1(x_1,y_1,z_1)=\pm {\cal U}W_1(x_1,y_1,z_1)$
and ${\cal U}W_2(x_2,y_2,x_2)=\pm {\cal U}W_2(x_2,y_2,x_2)$ hold simultaneously,
then Alice's goal can be deterministically reached via Bob and Charlie's collaboration.
Hence, as mentioned before, the success probability is fully determined by
the properties of ${\cal U}$ and $W_1(x_1,y_1,z_1)$ as well as $W_2(x_2,y_2,x_2)$,
particularly their mutual relations. After our intensive investigations,
we have found that in the following four cases the success probability
can be doubled or increased to unit, provided that some information
on the operation ${\cal U}$ is partially known in priori.

\vskip 0.2cm

{\bf (1)} \ $x_1=0$ and $y_1=-\frac{1}{\sqrt{2}}e^{i\phi_{1}}$ ($\phi_1$ is arbitrarily real).

In this case, one can easily get the two variant measuring bases
and the two $W$ operations, i.e.,
\begin{eqnarray}
|\xi_{{\bf 1}1}\rangle &=& e^{i(\tau_{1}+\phi_1)}|\vec{V}(0, -1/\sqrt{2}, 1/\sqrt{2})\rangle, \\
|\xi_{{\bf 1}2}\rangle &=&  e^{i\tau_{2}}|\vec{V}(- 2/\sqrt{6}, 1/\sqrt{6}, 1/\sqrt{6})\rangle, \\
W_{{\bf 1}1} &=&  e^{-i(\tau_{1}+\phi_1)}(-|1\rangle\langle1| + |2\rangle\langle2|)/\sqrt{2}, \\
W_{{\bf 1}2} &=& e^{-i\tau_{2}}(-2|0\rangle\langle0|+|1\rangle\langle1|+|2\rangle\langle2|)/\sqrt{6}.
\end{eqnarray}
If ${\cal U}W_{{\bf 1}1}= W_{{\bf 1}1} {\cal U}$ or ${\cal U}W_{{\bf 1}1}=- W_{{\bf 1}1}{\cal U}$,
then ${\cal U}$ should take one of the following two forms
\begin{equation}
{\cal U}^{(1)}=
\left(
\begin{array}{ccc}
e^{i\mu_{11}} &  0& 0 \\
  0& e^{i\mu_{12}}&  0 \\
 0 & 0 & e^{i\mu_{13} }
\end{array}
\right), \ \ \ \
{\cal U}^{(2)}=
\left(
\begin{array}{ccc}
e^{i\mu_{21}} &  0& 0 \\
  0& 0& e^{i\mu_{22}}  \\
 0 &e^{i\mu_{23}}& 0
\end{array}
\right),
\end{equation}
where $\mu$'s are arbitrarily real (same hereafter). This means that
if Bob measures $|\xi_{{\bf 1}1}\rangle_{b'}$  and  ${\cal U}$
belongs to either of the two restricted sets above,
Charlie can finally achieve Alice's goal.

Similarly, if ${\cal U} W_{{\bf 1}2}= W_{{\bf 1}2}{\cal U}$,
then ${\cal U}$ should be any matrix of the following couple of sets
\begin{equation}
{\cal U}^{(3)}=
\left(
\begin{array}{ccc}
e^{i\mu_{31}}& 0 &  0 \\
0&\cos\mu_{35} e^{i(\mu_{32}+\mu_{34})}& \sin\mu_{35} e^{i(\mu_{33}+\mu_{34})} \\
  0 & -\sin\mu_{35} e^{-i(\mu_{33}-\mu_{34})} & \cos\mu_{35} e^{-i(\mu_{32}-\mu_{34})}
\end{array}
\right),
\end{equation}
\begin{equation}
{\cal U}^{(4)}=
\left(
\begin{array}{ccc}
e^{i\mu_{41}}& 0 &  0 \\
0&\cos\mu_{45} e^{i(\mu_{42}+\mu_{44})} & \sin\mu_{45} e^{i(\mu_{43}+\mu_{44})} \\
  0 & \sin\mu_{45} e^{-i(\mu_{43}-\mu_{44})} & -\cos\mu_{45} e^{-i(\mu_{42}-\mu_{44})}
\end{array}
\right).
\end{equation}
This implies that if Bob gets $|\xi_{{\bf 1}2}\rangle_{b'}$ via measurement
and ${\cal U}$ belongs to either ${\cal U}^{(3)}$ or ${\cal U}^{(4)}$,
Charlie can fulfill Alice's operation ${\cal U}$ on his qutrit $c$ in the end.
Incidentally, there does not exist any ${\cal U}$ which satisfies
${\cal U} W_{{\bf 1}2}= - W_{{\bf 1}2}{\cal U}$.

Let ${\cal U}^{(12)}={\cal U}^{(1)}\cup~ {\cal U}^{(2)}$ and
${\cal ~U}^{(34)}= {\cal U}^{(3)}\cup{\cal U}^{(4)}$.
Surely, if ${\cal U} \in [{\cal U}^{(12)}\cap{\cal U}^{(34)}]$,
then the success probability of the scheme is unit when taking
account of the success probability 1/3 with the measured $|\xi_{0}\rangle_{b'}$.
In this case, one is readily to see that ${\cal U}^{(12)}\cap{\cal U}^{(34)}={\cal U}^{(12)}$.
In contrast, in the case that
${\cal U} \in [{\cal U}^{(34\setminus 12)}={\cal U}^{(34)}\setminus {\cal U}^{(12)}]$,
then the total success probability is 2/3.

\vskip 0.2cm
{\bf (2)}\ \ $x_{1}=-\frac{1}{\sqrt{2}}e^{i\phi_{2}}$ and $y_{1}=0$.

Two variant measuring bases and two $W$ operations corresponding to this case are
\begin{eqnarray}
&&|\xi_{{\bf 2}1}\rangle=  e^{i(\tau_{1}+\phi_2)}|\vec{V}(-1/\sqrt{2},0,1/\sqrt{2})\rangle,\\
&&|\xi_{{\bf 2}2}\rangle= e^{i\tau_{2}}|\vec{V}(1/\sqrt{6}, -2/\sqrt{6},1/\sqrt{6})\rangle,\\
&&W_{{\bf 2}1}=  e^{-i(\tau_{1}+\phi_2)}(-|0\rangle\langle0| +|2\rangle\langle2|)/\sqrt{2},\\
&&W_{{\bf 2}2}=e^{-i\tau_{2}}(|0\rangle\langle0|-2|1\rangle\langle1|+1|2\rangle\langle2|)/\sqrt{6}.
\end{eqnarray}
If ${\cal U} \in {\cal U}^{(15)}$, where ${\cal U}^{(15)}={\cal U}^{(1)}\cup~{\cal U}^{(5)}$ and
\begin{equation}
{\cal U}^{(5)}=
\left(
\begin{array}{ccc}
0 &  0& e^{i\mu_{51}} \\
  0& e^{i\mu_{52}}&  0 \\
e^{i\mu_{53}} & 0 & 0
\end{array}
\right),
\end{equation}
then ${\cal U}W_{{\bf 2}1}= W_{{\bf 2}1} {\cal U}$ or ${\cal U}W_{{\bf 2}1}= - W_{{\bf 2}1} {\cal U}$.
It means that if Bob's measurement result is $|\xi_{{\bf 2}1}\rangle_{b'}$
and ${\cal U}$ belongs the restricted set ${\cal U}^{(15)}$,
Charlie can finally reconstruct the state ${\cal U}|P\rangle$ on his qutrit $c$,
as implies that with the two agents' help with the shared entanglement and LOCC
Alice's operation ${\cal U}$ has been successfully executed on the remote qutrit $c$
in Charlie's site. Of course, such circumstance only appears with probability 1/3.

Similarly, if ${\cal U} \in {\cal U}^{(67)}$,
where ${\cal U}^{(67)}= {\cal U}^{(6)}\cup{\cal U}^{(7)}$ and
\begin{equation}
{\cal U}^{(6)}=
\left(
\begin{array}{ccc}
  \cos\mu_{65} e^{i(\mu_{62}+\mu_{64})}             & 0 &  \sin\mu_{65} e^{i(\mu_{63}+\mu_{64})} \\
0 &e^{i\mu_{61}}& 0 \\
  -\sin\mu_{65} e^{-i(\mu_{63}-\mu_{64})}&0  & \cos\mu_{65} e^{-i(\mu_{62}-\mu_{64})}
\end{array}
\right),
\end{equation}
\begin{equation}
{\cal U}^{(7)}=
\left(
\begin{array}{ccc}
\cos\mu_{75} e^{i(\mu_{72}+\mu_{74})} &     0                                  &  \sin\mu_{75} e^{i(\mu_{73}+\mu_{74})} \\
0                                    &  e^{i\mu_{71}} & 0 \\
\sin\mu_{75} e^{-i(\mu_{73}-\mu_{74})}                   & 0 & -\cos\mu_{75} e^{-i(\mu_{72}-\mu_{74})}
\end{array}
\right),
\end{equation}
then ${\cal U}W_{{\bf 2}2}= \pm W_{{\bf 2}2} {\cal U}$.
Hence in the case, ${\cal U}$ can be finally performed on the qutrit $c$
in a deterministic manner provided that Bob measures $|\xi_{{\bf 2}2}\rangle_{b'}$
and notifies him of the result.
Surely, if ${\cal U} \in [{\cal U}^{(15)}\cap{\cal U}^{(67)}={\cal U}^{(15)}]$,
then the success probability of the scheme is unit. Apparently, in the case that
${\cal U} \in [{\cal U}^{(67\setminus 15)}={\cal U}^{(67)}\setminus {\cal U}^{(15)}]$,
then the total success probability is still 2/3.
At last, it is worthy mentioning that {\bf (1)} is completely same as {\bf (2)}
if $|0\rangle$ exchanges with $|1\rangle$.

\vskip 0.2cm
{\bf (3)}\ \ $x_{1}=-y_{1}=\frac{1}{\sqrt{2}}e^{i\phi_{3}}$.

The two variant measuring bases and two $W$ operations corresponding to the conditions above are
\begin{eqnarray}
&&|\xi_{{\bf 3}1}\rangle =e^{i(\tau_{1}+\phi_3)}|\vec{V}(1/\sqrt{2}, -1/\sqrt{2}, 0)\rangle,\\
&&|\xi_{{\bf 3}2}\rangle=e^{i\tau_{2}}|\vec{V}(1/\sqrt{6}, 1/\sqrt{6}, -2/\sqrt{6})\rangle,\\
&&W_{{\bf 3}1}= e^{-i(\tau_{1}+\phi_3)}(-|0\rangle\langle0| + |1\rangle\langle1|)/\sqrt{2},\\
&&W_{{\bf 3}2}=e^{-i\tau_{2}}(|0\rangle\langle0|+|1\rangle\langle1|-2|2\rangle\langle2|)/{\sqrt{6}}.
\end{eqnarray}
If ${\cal U}W_{{\bf 3}1}=\pm W_{{\bf 3}1}{\cal U}$, then ${\cal U} \in {\cal U}^{(18)}$,
where ${\cal U}^{(18)}={\cal U}^{(1)}\cup {\cal U}^{(8)}$ and
\begin{equation}
{\cal U}^{(8)}=
\left(
\begin{array}{ccc}
0 & e^{i\mu_{81}}& 0 \\
 e^{i\mu_{82}}&0 &  0 \\
 0 & 0 & e^{i\mu_{83} }
\end{array}
\right).
\end{equation}
If ${\cal U}W_{{\bf 3}2}=W_{{\bf 3}2}{\cal U}$,
then ${\cal U} \in {\cal U}^{(910)}$,
where ${\cal U}^{(910)}={\cal U}^{(9)}\cup {\cal U}^{(10)}$ and
\begin{equation}
{\cal U}^{(9)}=
\left(
\begin{array}{ccc}
  \cos\mu_{95} e^{i(\mu_{92}+\mu_{94})}             & \sin\mu_{95} e^{i(\mu_{93}+\mu_{94})}&  0 \\
-\sin\mu_{95} e^{-i(\mu_{93}-\mu_{94})}&       \cos\mu_{95} e^{-i(\mu_{92}-\mu_{94})}           & 0 \\
 0 &0  & e^{i\mu_{91}}
\end{array}
\right),
\end{equation}
\begin{equation}
{\cal U}^{(10)}=
\left(
\begin{array}{ccc}
  \cos\mu_{105} e^{i(\mu_{102}+\mu_{104})}             & \sin\mu_{105} e^{i(\mu_{103}+\mu_{104})}&  0 \\
\sin\mu_{105} e^{-i(\mu_{103}-\mu_{104})}&       -\cos\mu_{105} e^{-i(\mu_{102}-\mu_{104})}           & 0 \\
 0 &0  & e^{i\mu_{101}}
\end{array}
\right).
\end{equation}
In addition, after extensive investigations it is found that
${\cal U}W_{{\bf 3}2}=-W_{{\bf 3}2}{\cal U}$ does not hold for any ${\cal U}$.
Consequently,  when ${\cal U} \in [{\cal U}^{(18)}\cap{\cal U}^{(910)}={\cal U}^{(18)}]$,
the scheme success probability can reach 1. On the contrary,
if ${\cal U} \in [{\cal U}^{(910\setminus 18)}={\cal U}^{(910)}\setminus {\cal U}^{(18)}]$,
the success probability of the scheme is 2/3. Obviously, one can see that
the item {\bf (3)} is as same as the item {\bf (2)} [or the item {\bf (1)}]
provided that $|2\rangle$ is exchanged with $|1\rangle$ [or $|0\rangle$].

\vskip 0.2cm
{\bf (4)} \ \ $x_1\neq 0$, $y_1\neq 0$ and $x_1\neq-y_1$.

With these conditions, one gets the two variant measuring bases  and the two $W$ operations
\begin{eqnarray}
&&|\xi_{{\bf 4}1}\rangle = e^{i\tau_{1}}|{\vec{V}} (x_{1},y_{1}, -x_{1}- y_{1})\rangle,  \\
&&|\xi_{{\bf 4}2}\rangle = e^{i\tau_{2}}|{\vec{V}}(x_2,y_2,z_2)\rangle, \\
&&W_{{\bf 4}1} = e^{-i\tau_{1}}(x_{1}^{\ast}|0\rangle\langle0| +y_{1}^{\ast}|1\rangle\langle1|+(-x_{1}^{\ast}-y_{1}^{\ast})|2\rangle\langle2|),\\
&&W_{{\bf 4}2} = e^{-i\tau_{2}}(x_2^{\ast}|0\rangle\langle0| + y_2^{\ast}|1\rangle\langle1|+z_2^{\ast}|2\rangle\langle2|).
\end{eqnarray}
After extensive studies, we have found that no ${\cal U}$ exists to satisfy
${\cal U}W_{{\bf 4}1}= - W_{{\bf 4}1}{\cal U}$ or
${\cal U}W_{{\bf 4}2}=-W_{{\bf 4}2}{\cal U}$.
Instead, if ${\cal U}W_{{\bf 4}1}=W_{{\bf 4}1}{\cal U}$
or ${\cal U}W_{{\bf 4}2}=W_{{\bf 4}2}{\cal U}$ should hold,
then one can find that ${\cal U}$ must be in ${\cal U}^{(1)}$.
Alternatively, if ${\cal U} \in {\cal U}^{(1)}$,
both $W_{{\bf 4}1}$ and $W_{{\bf 4}2}$ commute with it.
Nevertheless, as stressed before, $x_1$ and $y_1$ are not
independent but constrained by a relation. Because of this,
not all measuring bases can be used to finally fulfill Alice's task.
In other words, only some specific bases are applicable, including those
occurring in items {\bf (1-3)}. In reality, one can conveniently
take different measuring bases in terms of the feasible state discrimination ability.
Here we pick out another two discrete examples to demonstrate that
other measuring bases are also applicable:
\begin{eqnarray}
&&|\xi'_{{\bf 4}1}\rangle=e^{i\tau_{1}} |\vec {V}(1/\sqrt{3},e^{\frac{2\pi}{3}i}/\sqrt{3},e^{\frac{4\pi}{3}i}/\sqrt{3})\rangle,  \\
&&|\xi'_{{\bf 4}2}\rangle=e^{i\tau_{2}}|\vec{V}(1/\sqrt{3},e^{\frac{4\pi}{3}i}/\sqrt{3},e^{\frac{2\pi}{3}i}/\sqrt{3})\rangle, \\
&&W'_{{\bf 4}1}=e^{-i\tau_{1}}(|0\rangle\langle0|+e^{\frac{4\pi}{3}i}|1\rangle\langle1| +e^{\frac{2\pi}{3}i}|2\rangle\langle2|)/\sqrt{3},\\
&&W'_{{\bf 4}2}=e^{-i\tau_{2}}(|0\rangle\langle0|+e^{\frac{2\pi}{3}i}|1\rangle\langle1| +e^{\frac{4\pi}{3}i}|2\rangle\langle2|)/\sqrt{3}. \\
\nonumber \\
&&|\xi''_{{\bf 4}1}\rangle=e^{i\tau_{1}}|{\vec{V}} [(i/2, 1/2,-(1+i)/2)]\rangle,  \\
&&|\xi''_{{\bf 4}2}\rangle=e^{i\tau_{2}}|{\vec{V}}[(-2\sqrt{3}+\sqrt{3}i)/6, (-2\sqrt{3}i+\sqrt{3})/6 ,(\sqrt{3}+\sqrt{3}i)/6]\rangle, \\
&&W''_{{\bf 4}1}=e^{-i\tau_{1}}(-i|0\rangle\langle0|+|1\rangle\langle1|+ (-1+i)|2\rangle\langle2|)/2,\\
&&W''_{{\bf 4}2}=e^{-i\tau_{2}} [(-2\sqrt{3}-\sqrt{3}i)|0\rangle\langle0|+(2\sqrt{3}i+\sqrt{3})|1\rangle\langle1| +(\sqrt{3}-\sqrt{3}i)|2\rangle\langle2|]/6.
\end{eqnarray}

\vskip 0.3cm
\noindent {\bf 4 Comparisons and discussions }

Now let us move to compare our schemes from the following four aspects:
the resource consumption consisting of its classical and quantum parts,
the necessary-operation complexity including its difficulty and intensity,
the scheme success probability and the intrinsic efficiency of the scheme.
We have already summarized the two schemes in Table 1 with respect to
the four aspects. The intrinsic efficiency of any single-qutrit operation
sharing scheme reads
\begin{eqnarray}
\eta=\frac{P}{Q_{t}+C_t},
\end{eqnarray}
where $Q_{t}$ is the number of the qutrits which are used as quantum channels
(except for those chosen for security checking), $C_t$ is the classical trits
transmitted and $P$ is the final success probability.

\vskip 0.6cm
\begin{center}
\begin{minipage}{400pt}
Table 1. Comparisons between our two schemes S1 and S2. QRC: quantum resource consumption;
NO: necessary operations; CRC: classical resource consumption; GB: generalized Bell state;
GG: generalized GHZ state; GM: generalized Bell-state measurement; SM: single-qutrit measurement;
SO: single-qutrit unitary operation. Note that, in the text the two single-qutrit operations on
$c$ are used for the sake of convenient expressions. Because they can be commuted with other operations
on other qutrits, they can be incorporated as one.
\end{minipage}
\vskip 0.2cm
\begin{tabular*}{12cm}{@{\extracolsep{\fill}}ccccccc}
\hline
S   & ${\cal U}$                                   & QRC            &  NO                                      & CRC         &P        & $\eta$  \\ \hline
S1  &arbitrary                                     &  GB, GG        &  2 GMs, SM, 2 SOs                        & 5 ctrits    & 1       & 1/10    \\
S2  &arbitrary                                     & 2 GBs          &  ${\cal V}$,  GM, 2 SMs, 3 SOs           & 4 ctrits    & 1/3     & 1/24    \\
S2  &${\cal U}^{(34\setminus 12)}$                 & 2 GBs          &  ${\cal V}$,  GM, 2 SMs, 3 SOs           & 4 ctrits    & 2/3     & 1/12    \\
S2  &${\cal U}^{(67\setminus15)}$                  & 2 GBs          &  ${\cal V}$,  GM, 2 SMs, 3 SOs           & 4 ctrits    & 2/3     & 1/12    \\
S2  &${\cal U}^{(910\setminus 18)}$                & 2 GBs          &  ${\cal V}$,  GM, 2 SMs, 3 SOs           & 4 ctrits    & 2/3     & 1/12    \\
S2  &${\cal U}^{(12)}$                             & 2 GBs          &  ${\cal V}$,  GM, 2 SMs, 3 SOs           & 4 ctrits    & 1       & 1/8     \\
S2  &${\cal U}^{(15)}$                             & 2 GBs          &  ${\cal V}$,  GM, 2 SMs, 3 SOs           & 4 ctrits    & 1       & 1/8     \\
S2  &${\cal U}^{(18)}$                             & 2 GBs          &  ${\cal V}$,  GM, 2 SMs, 3 SOs           & 4 ctrits    & 1       & 1/8     \\
\hline
\end{tabular*}\\
\end{center}

\vskip 0.5cm

Obviously, one can see that the general scheme (i.e., S1) is a deterministic one,
that is, Alice's task can be achieved with unit probability. As can be seen from
the first line of the table. As a matter of fact, the first scheme is essentially
an ordering incorporation of three processes, i.e., QST and Alice's operation
performance as well as QSTS. Although it looks like a trivial one, it actually
offers an upper limit of resource consumptions and shows the complexity of necessary
operations in accomplishing the quantum task. Specifically, in the scheme two units
of entanglements and five trits of the classical communication cost are indispensable,
the necessary operations are two generalized Bell-state measurements and two
single-qutrit unitary operations as well as a single-qutrit measurement,
the intrinsic efficiency of the scheme is 1/10. The distinct feature of this scheme
is its universality for sharing any single-qutrit operation in a deterministic manner.
All these can be taken as a useful frame of reference for some other optimal schemes
which might be proposed later aiming at some special considerations.

In our second scheme, the quantum channels are changed by instituting the generalized
GHZ state with the generalized Bell state. Very intuitively, the quantum resource consumption
is decreased. Moreover, the classical resource consumption is also reduced, too.
All these can be seen from the table by inspecting the third and fifth columns of
the the schemes S1 and S2. Nonetheless, the second scheme can not be simply
decomposed as the three processes mentioned in the last paragraph anymore.
This essential change will bring some variance, as will be see later.

As far as the sharing of any single-qutrit operation is concerned,
besides the common advantages mentioned in the last paragraph,
the quantum operation complexity in the second scheme is also apparently simplified.
Note that, generally speaking, the complexity of a generalized Bell-state
measurement can be decomposed into a series of ordering two-qutrit control operation
and a single single-qutrit operation as well as two single-qutrit measurements. Hence
its implementation difficulty is approximately equal to that of the operation ${\cal V}$
and a single-qutrit operation as well as two single-qutrit measurements.
All these advantages can be seen
from the second line contrasting to the first one in the table.
However, the cost of these resultant advantages is also very clear. It is obvious
that in the second scheme both the scheme success probability and its
intrinsic efficiency are much smaller than those in the first one.
Evidentally, these two indicators mentioned just are violently decreased.
The scheme S2 becomes a probabilistic one.

As for sharing the restricted sets of operations, two cases in the second scheme are grouped.
One considers the restricted sets listed in the second column of the third,
fourth and fifth lines in the table, i.e., ${\cal U}^{(34\setminus 12)}$,
${\cal U}^{(67\setminus15)}$ and ${\cal U}^{(910\setminus 18)}$. The other
treats the restricted sets shown in the second column of the last three
in the table, namely, ${\cal U}^{(12)}$, ${\cal U}^{(15)}$ and ${\cal U}^{(18)}$.
With a priori knowledge on these sets (only on the sets themselves, not the detailed
information of the elements in the sets), contrasting to the sharing an arbitrary operation,
one can easily find that in the former case both the scheme success probability
and the intrinsic efficiency are doubled while in the latter the two indicators are tripled.
We have mentioned before that the scheme S2 has become a probabilistic one
after the quantum channel change. In the present former case on restricted sets,
The scheme remains probabilistic. However, in the latter case, an essential variance happens,
i.e., the scheme has already changed to be a deterministic one. Alternatively,
the scheme success probability has been increased to 1. In this situation,
it is intriguing to compare the specific scheme S2 in this case with the general scheme S1.
They both have the unit success probability. Nonetheless, with the priori knowledge
on the restricted sets, S2 completely overwhelms S1 in all the four aspects
of quantum resource consumption, classical resource consumption, difficulty and
intensity of necessary operations, and the intrinsic efficiency. In this sense,
one can think that S2 is more optimal with the precondition of the priori knowledge.

At last, we want to briefly point out that the present schemes are actually the generalization of
the ZC schemes from the aspect of particle degrees. The former is a qutrit one while the latter
is a qubit one. In the ZC scheme on restricted sets, there only exist two kinds of success probabilities  (1/2 or 1),
and the success probability 1/2 corresponds to the universal applicability while the unit probability
to the successful application to two restricted sets of unitary operations. After the degree extension,
the results become more abundant. Easily one can see that more restricted sets occur and more possibilities
appear. Our present qutrit schemes actually contains the ZC schemes. Alternatively,
the present schemes can be reduced to the ZC schemes, as can be easily seen from
the restricted sets. In the case of the unit success probability, if one degree is frozen,
the restricted sets ${\cal U}^{(12)}$, ${\cal U}^{(15)}$ and ${\cal U}^{(18)}$
can be easily reduced to the diagonal or anti-diagonal sets in the ZC scheme. \\

\noindent {\bf 5 Summary}

To summarize, in this paper we have actually considered the shared remote control
in the form of quantum operation sharing on a single-qutrit state. By integrating
the ideas of quantum operation teleportation and quantum secret sharing, we have
presented two possible schemes. The first scheme is universally applicable for
any arbitrary single-qutrit unitary operation. The second scheme is generally a
probabilistic one. However, after intensive investigations we have found that,
if the operation ${\cal U}$ in question is known to belong to some restricted sets,
both the scheme success probability and its efficiency can be doubled or even tripled.
We have concretely compared the schemes in different cases from the four aspects of
quantum and classical resource consumption, necessary-operation complexity,
success probability and efficiency. In the tripled case the latter scheme becomes
a deterministic one and is more optimal than the general scheme. \\

\noindent {\bf Acknowledgements}

This work is supported by the Specialized Research Fund for the Doctoral Program of
Higher Education under Grant No.20103401110007, the NNSFC under Grant
Nos.10874122, 10975001, 51072002 and 51272003, the Program for Excellent
Talents at the University of Guangdong province (Guangdong Teacher Letter
[1010] No.79), and the 211 Project of Anhui University.\\

\noindent {\bf References}

\noindent[1] Ekert, A.K.
Phys. Rev. Lett. {\bf 67}, 661 (1991)

\noindent[2] Bennett, C.H., Brassard, G., Cr$\acute{e}$peau, C. Phys. Rev. Lett. {\bf 70}, 1895 (1993)

\noindent[3] Bouwmeester, D., Pan, J.W., Mattle, K., Eibl, M., Weinfurter, H., Zeilinger, A.
Nature {\bf 390}, 575 (1997)

\noindent[4] Deng, F.G., Long, G.L., Liu, X.S.
Phys. Rev. A  {\bf 68}, 042317 {2003}

\noindent[5] Cheung, C.Y., Zhang, Z.J.
Phys. Rev. A {\bf 80}, 022327 (2009)

\noindent[6] Lo, H.K.
Phys. Rev. A {\bf 62}, 012313 (2000)

\noindent[7] Huelga, S.F., Vaccaro, J.A., Chefles, A.
Phys. Rev. A {\bf 63}, 042303 (2001)

\noindent[8] Huelga, S.F., Plenio, M.B., Vaccaro, J.A.
Phys. Rev. A {\bf 65}, 042316 (2002)

\noindent[9] Hillery, M., Buzk, V., Berthiaume, A.
Phys. Rev. A {\bf 59}, 1829 (1999)

\noindent[10] Cleve, R., Gottesman, D., Lo, H.K.
Phys. Rev. Lett. {\bf 82}, 648 (1999)

\noindent[11] Lance, A.M., Symul, T., Bowen, W.P., Sanders, B.C., Lam, P.K.
Phys. Rev. Lett. {\bf 92}, 177903 (2005)

\noindent[12] Zhang, Z.J., Man, Z.X., Li, Y.:
Phys. Rev. A {\bf 71}, 044301 (2005)

\noindent[13] Zhang, Z.J., Man, Z.X.
Phys. Rev. A  {\bf 72}, 022303 (2005)

\noindent[14] Deng, F.G., Li, X.H., Zhou, H.Y., Zhang, Z.J.
Phys. Rev. A {\bf 72}, 044302 (2005)

\noindent[15] Gaertner, S., Kurtsiefer, C., Bourennane, M., Weinfurter, H.
Phys. Rev. Lett. {\bf 98}, 020503 (2007)

\noindent[16] Muralidharan, S., Panigrahi, P.K.
Phys. Rev. A {\bf 77}, 032321  (2008)

\noindent[17] Muralidharan, S., Panigrahi, P.K.
Phys. Rev. A {\bf 78}, 062333 (2008)

\noindent[18] Muralidharan, S., Jain, S., Panigrahi, P.K.
Opt. Commun. {\bf 284}, 1082 (2011)

\noindent[19] Choudhury, S., Muralidharan, S., Panigrahi, P.K.
J. Phys. A {\bf 42}, 115303 (2009)

\noindent[20] Saha, D., Panigrahi, P.K.
Quantum Inf. Process. {\bf 11}, 615 (2012)

\noindent[21] Deng, F.G., Long, G.L.
Phys. Rev. A {\bf 69},  052319 (2004)

\noindent[22] Zhang, Z.J., Liu, J., Wang, D., Shi, S.H.
Phys. Rev. A {\bf 75}, 026301 (2007)

\noindent[23] Zhu, A.D., Xia, Y., Fan, Q.B., Zhang, S.
Phys. Rev. A {\bf 73}, 022338 (2006)

\noindent[24] Nielsen, M.A.
Phys. Rev. Lett. {\bf 93}, 040503 (2004)

\noindent[25] Briegel, H.J., Raussendorf, R.
Phys. Rev. Lett. {\bf 86}, 910 (2001)

\noindent[26] Zhang, Z.J., Cheung, C.Y.
J. Phys. B {\bf 44}, 165508 (2011)

\noindent[27] Ye, B.L., Liu, Y.M., Liu, X.S., Zhang, Z.J.
Chin. Phys. Lett. {\bf 30}, 020301 (2013)

\noindent[28] Ji, Q.B., Liu, Y.M., Yin, X.F., Liu, X.S. Zhang, Z.J.
Quantum Inf. Process. (in press)

\noindent[29] Wang, S.F., Liu, Y.M., Chen, J.L., Liu, X.S., Zhang, Z.J.
Quantum Inf. Process. (in press)

\noindent[30] Giampaolo, S.M., Illuminati, F.
Phys. Rev. A {\bf 76}, 042301 (2007)

\noindent[31] P\'{e}rez, A.
Phys. Rev. A {\bf 81}, 052326 (2010)

\noindent[32] Zhou, J.D., Hou, G., Zhang, Y.D.
Phys. Rev. A {\bf 64}, 012301 (2001)

\noindent[33] Zeng, B., Zhang, P.
Phys. Rev. A {\bf 65}, 022316 (2002)

\noindent[34] Yu, C.S., Song, H.S., Wang, Y.H.
Phys. Rev. A {\bf 73}, 022340 (2006)

\noindent[35] Xia, Y., Song, H.S.
Phys. Lett. A {\bf 364}, 117  (2007)

\noindent[36] Wei, H.R., Ren, B.C., Deng, F.G.
Quantum Inf. Process.  {\bf 12}, 1109  (2013)

\noindent[37] Bogdanski, J., Rafiei, N., Bourennane, M.
Phys. Rev. A {\bf 78}, 062307  (2008)

\noindent[38] Zhang, Z.J., Liu, Y.M., Fang, M.
Int. J. Theor. Phys. {\bf 18}, 1885 (2007)

\noindent[39] Wang, C., Deng, F.G., Li, Y.S., Liu, X.S., Long, G.L.
Phys. Rev. A {\bf 71} 044305  (2005)

\end{document}